# Study of Brain Connectivity by Multichannel EEG Quaternion Principal Component Analysis for Alzheimer's Disease Classification

Kevin Hung*, Gary Man-Tat Man, and Jincheng Wang

School of Science and Technology, Hong Kong Metropolitan University, Ho Man Tin, Kowloon, Hong Kong
{khung, mtman}@hkmu.edu.hk

* Corresponding Author: Kevin Hung, khung@hkmu.edu.hk

* Regular Paper

***Abstract***: The early detection of Alzheimer's disease (AD) through widespread screening has emerged as a primary strategy to mitigate the significant global impact of AD. EEG measurements offer a promising solution for extensive AD detection. However, the intricate and nonlinear dynamics of multichannel EEG signals pose a considerable challenge for real-time AD diagnosis. This paper introduces a novel algorithm, which is based on Quaternion Principal Component Analysis (QPCA) of multichannel EEG signals, for AD classification. The algorithm extracts high dimensional correlations among different channels to generate features that are maximally representative with minimal information redundancy. This provides a multidimensional and precise measure of brain connectivity in disease assessment. Simulations have been conducted to evaluate the performance and to identify the most critical EEG channels or brain regions for AD classification. The results reveal a significant drop of connectivity measure in the alpha bands. The average AD classification accuracy for all 4-channel combinations reached 95%, while some particular permutations of channels achieved 100% accuracy rate. Furthermore, the temporal lobe emerges as one of the most important regions in AD classification given that the EEG signals are recorded during the presentation of an auditory stimulant. The selection of key parameters of the QPCA algorithm have been evaluated and some recommendations are proposed for further performance enhancement. This paper marks the first application of the QPCA algorithm for AD classification and brain connectivity analysis using multichannel EEG signals.

## 1. Introduction

Alzheimer's disease (AD) is a neurodegenerative condition that progressively impairs cognitive abilities, affecting memory, thinking, and the capacity to carry out daily tasks. AD accounts for 60-70% of dementia cases. The World Health Organization reported that over 55 million people worldwide were living with dementia, with approximately 10 million new cases diagnosed annually. The increasing prevalence of AD places a significant burden on global healthcare and social systems [1]. Regrettably, there is currently no effective cure for AD. Existing treatments can only temporarily slow symptom progression [2]. Moreover, AD may take decades to progress from the onset of pathological changes to the manifestation of clinical symptoms, underscoring the critical importance of early detection and widespread disease screening [3]. Various methods are used for early AD detection. These include neuroimaging biomarkers based on magnetic resonance imaging (MRI) and positron emission tomography (PET), as well as biochemical biomarkers that are based on cerebrospinal fluid (CSF) [4]. However, neuroimaging is costly and not widely accessible, and CSF testing is invasive. These methods are not suitable for broad AD diagnosis, particularly in low and middle-income countries.

### 1.1. Overview of EEG Processing

Electroencephalography (EEG) is a non-invasive and relatively cost-effective method for evaluating neurophysiological functions. It involves measuring the electrical signals generated by active neurons in the brain using electrodes placed on the scalp. These signals, which reflect the dynamics and synchronization of brain activity across different cortical regions, have led to the widespread use of EEG in diagnosing and assessing the progression of neurological and psychiatric disorders, such as epilepsy [5], depression [6, 7], dementia, and schizophrenia [8, 9]. However, analyzing EEG data can be challenging due to its complexity and nonlinear dynamics [10]. EEG measurements comprise real-time multichannel signals, typically from 19 up to 256 channels, each probing different areas of the brain. Developing algorithms to analyze such vast, interconnected, and temporal data is highly complex. A considerable amount of research is being dedicated to enhancing EEG signal processing for identifying brain-related diseases, with emphasis on multichannel EEG signal processing and brain connectivity analysis.

Multichannel EEG signals can be analyzed using various methods, including time-domain analysis [11], frequency-domain analysis [11], wavelet analysis [12], independent component analysis (ICA) [13], and principal component analysis (PCA) [14, 15]. Among these, PCA is a widely used technique for reducing dimensionality and extracting relevant features from EEG signals. H. Rajaguru et al. utilized PCA to simplify the dimension of complex multichannel EEG signals and employed the Adaboost classifier to enhance the performance of epileptic classification [14]. Similarly, G. Biagetti et al. used a PCA-based feature extraction method for AD classification from EEG signals and compared the recognition accuracy among several machine learning algorithms. Their results indicated that the support vector machine (SVM) had a significantly lower computational cost and higher accuracy compared to other classifiers [15].

The EEG-based analysis of brain connectivity, which quantifies the statistical relationship between electrode pairs or brain regions, is also found useful primarily because AD is considered a disconnection syndrome due to its destructive characteristics [16, 17]. Numerous measures have been proposed to quantify functional connectivity, and most studies have reported a loss of functional connectivity in early AD, especially in the EEG's alpha band [8, 9].

## 1.2. Limitations of Conventional EEG Processing

While previous studies in conventional multichannel EEG processing have provided valuable insights, certain limitations impede their overall effectiveness in identifying brain-related diseases. In PCA-based multichannel EEG processing, the effectiveness of PCA relies on the assumption that the data is embedded in an approximately linear low-dimensional space [18]. PCA primarily retains the second-order matrix information in the original data, but the variance does not fully represent the amount of information required for classification [19]. The conventional method of merging multichannel EEG features also overlooks the interrelations among the channels, leading to information loss in classification [20]. On the other hand, the overall effectiveness of traditional functional connectivity measures is compromised by the assumption of temporal stationarity and the information loss due to pairwise correlation.

### 1.2.1 Temporal Stationarity

Functional connectivity measures generally assume that the functional connections in the brain remain constant over a task or data collection period. However, brain activity is highly context-sensitive and activity-dependent, with intrinsic brain activity being dynamic and fluctuating over time. This static connectivity measure may not fully capture the changes in brain connectivity. Recently, some researchers have explored dynamic functional connectivity (dFC) by using a sliding-window technique to study the patterns in these temporal connections among brain regions [21, 22]. Because most dFC studies were conducted with MRI recordings, the study of dFC based on EEG recordings is of significant interest [23].

### 1.2.2 Pairwise Correlation

All functional connectivity measures are based on pairwise correlation or the statistical relationship between electrode pairs or brain regions. Given the complexity of brain activities and the high correlation between electrodes at different regions in functional connectivity, pairwise correlation can result in the loss of high-order correlation of multiple electrodes and regions in the brain. To mitigate such information loss, S. Panwar et al. recently proposed a recursive method to compute high-order correlation among three electrodes at a time in order to identify some dominant connectivity patterns [24]. However, the study of multichannel relations among dynamic brain networks is still in at preliminary stage.

The limitations of the existing methods for handling multichannel EEG signals in disease classification include several key issues. Firstly, these methods often fail to capture the interrelation among different EEG channels, which weakens their discriminative ability in disease classification. Secondly, many methods rely on static functional connectivity networks, which do not fully capture the brain's complex and dynamic nature. Lastly, traditional functional connectivity networks are constructed using pairwise correlations, leading to the loss of high-order information that could be crucial for understanding complex brain interactions.

## 1.3. Quaternion-based Multichannel EEG Processing

To overcome the abovementioned limitations, better ways to represent and manipulate with multichannel EEG signals are required to extract higher dimensional correlation among the channels and to consolidate these dynamic information over the whole recording. Recently, the advancements in multidimensional sensor technologies have underscored the need for signal processing algorithms in the quaternion domain, given its significant potential for modeling 4-D data and multichannel correlation. In various fields, including computer vision and graphics [25, 26], robotics and navigation systems [27], and vector-sensor signal processing [28], quaternion representation and computation methods have been successfully applied to process multichannel data and to extract new channel interrelated information. Similarly, EEG signals can be processed as hypercomplex channels, rather than just separate channels. P. Batres-Mendoza et al. have proposed a quaternion-based technique to model multichannel EEG signals as a single entity to detect and interpret cognitive activity in brain-computer interface (BCI) applications [29]. S. Enshaeifar et al. demonstrated how the augmented quaternion-valued singular spectrum analysis (QSSA) can be used to extract features reflecting the co-channel coupling of EEG signals in sleep analysis, even at very low signal-to-noise ratios [30]. Recently, W. Liu et al. showed that quaternion PCA (QPCA) improved accuracy in color face recognition applications [31]. Y. Zhao et al. proposed using quaternion representation to fuse multichannel EEG signals and applying QPCA to extract combined channel features for epilepsy detection [32]. However, the use of pure quaternion representation instead of a full 4-channel representation in the existing QPCA algorithm may lead to loss of mutual relation information. Also, QPCA was directly applied on raw EEG signals, which may consume heavy computational load and increase the difficulty in real-time implementation. Furthermore, the use of the singular value of the first few eigenvectors as principal features in the existing QPCA algorithm may weaken the generalization power for classification. Therefore, the existing QPCA algorithm needs to be further enhanced for extracting interrelated information among brain functional connectivity and for improved performance in brain-related disease detection.

### 1.4. Aim of Study

In this paper, an enhanced QPCA algorithm is proposed for processing multichannel EEG. The method aims to enhance early detection and classification of AD through EEG analysis, which can facilitate timely intervention. The algorithm provides an efficient way to address the limitations of traditional multichannel processing and functional connectivity measures based on EEG signals. It consolidates the temporal connectivity in the dynamics of brain activities and provides multichannel correlation by using quaternion representation and computations. The proposed algorithm will be evaluated with a SVM classifier on an EEG dataset of AD patients. The main contributions of this paper are summarized as follows: (i) To the best of our knowledge, this is the first application of QPCA in AD classification and brain connectivity analysis; (ii) Utilizing the new QPCA algorithm, a 4-D functional connectivity matrix is established across various brain frequency bands. The findings indicate a reduction in connectivity within the alpha band and an elevation in the delta and beta bands in AD patients with AD. These results are consistent with relevant research findings; (iii) Simulations are carried out for all permutations of 4-channel EEG combinations to evaluate the most informative channels and brain region for AD classification. The identified channels and region of interest (ROI) of the brain are consistent with those found in previous studies. The rest of this paper is organized as follows. Section 2 introduces the methodologies, including a detailed explanation of the enhanced QPCA algorithm, a description of the dataset, the operational flow, and the specific simulation settings. Section 3 showcases the results. Section 4 will delve into an analysis of the results and discuss future work. The conclusion is given in Section 5.

## 2. Methodologies

### 2.1 Quaternion Algebra

Quaternions are 4-D hypercomplex numbers. A quaternion $\boldsymbol{q} \in \mathbb{H}$ is defined as:

$$\boldsymbol{q} = w + x\boldsymbol{i} + y\boldsymbol{j} + z\boldsymbol{k} \qquad (1)$$

, where $w, x, y, z$ are all real numbers, and the imaginary units $(\boldsymbol{i}, \boldsymbol{j}, \boldsymbol{k})$ have the following properties:

$$\begin{aligned} \boldsymbol{ij} &= \boldsymbol{k} = -\boldsymbol{ji} \\ \boldsymbol{jk} &= \boldsymbol{i} = -\boldsymbol{kj} \\ \boldsymbol{ki} &= \boldsymbol{j} = -\boldsymbol{ik} \\ \boldsymbol{i}^2 &= \boldsymbol{j}^2 = \boldsymbol{k}^2 = \boldsymbol{ijk} = -\mathbf{1}. \end{aligned} \qquad (2)$$

Quaternions form a noncommutative normed division algebra $\mathbb{H}$, for $\boldsymbol{q_0}, \boldsymbol{q_1} \in \mathbb{H}$ the multiplication $\boldsymbol{q_0 q_1} \neq \boldsymbol{q_1 q_0}$, and is given by:

$$\boldsymbol{q_0 q_1} = \begin{aligned} &(w_0 w_1 - x_0 x_1 - y_0 y_1 - z_0 z_1) \\ &+ (w_0 x_1 + x_0 w_1 + y_0 z_1 - z_0 y_1)\boldsymbol{i} \\ &+ (w_0 y_1 - x_0 z_1 + y_0 w_1 + z_0 x_1)\boldsymbol{j} \\ &+ (w_0 z_1 + x_0 y_1 - y_0 x_1 + z_0 w_1)\boldsymbol{k}. \end{aligned} \qquad (3)$$

Let $\boldsymbol{Q} \in \mathbb{H}^{m \times n}$ be a quaternion matrix of rank $r$, the singular value decomposition (SVD) of $\boldsymbol{Q}$ is defined as follows [22]:

$$\boldsymbol{Q} = \boldsymbol{U} \begin{pmatrix} \Sigma_r^{1/2} & 0 \\ 0 & 0 \end{pmatrix} \boldsymbol{V}^H \qquad (4)$$

, where $\Sigma_r = \mathrm{diag}(\sigma_1, \dots, \sigma_r)$ is a real diagonal matrix and has $r$ non-null entries in descending order on its diagonal denoted the singular values of $\boldsymbol{Q}$. $\boldsymbol{U} \in \mathbb{H}^{n \times n}$, and $\boldsymbol{V} \in \mathbb{H}^{m \times m}$ are quaternion unitary matrices, contain respectively the left and right quaternion singular vectors of $\boldsymbol{Q}$, $\boldsymbol{V}^H$ is the conjugate transpose matrix of $\boldsymbol{V}$. Multichannel EEG signals can be represented in quaternion domain, denoted as $\boldsymbol{q}(t)$, as follow:

$$\boldsymbol{q}(t) = \boldsymbol{Ch_1}(t) + \boldsymbol{Ch_2}(t) \cdot \boldsymbol{i} + \boldsymbol{Ch_3}(t) \cdot \boldsymbol{j} + \boldsymbol{Ch_4}(t) \cdot \boldsymbol{k} \quad (5)$$

, where $\boldsymbol{Ch_1}(t), \boldsymbol{Ch_2}(t), \boldsymbol{Ch_3}(t)$, and $\boldsymbol{Ch_4}(t)$ are real value measured at sampling point $\mathbf{t}$ of the four channels of EEG signal being selected for analysis. Therefore, the four real-valued EEG channel time series data are converted to a more compact quaternion-valued time-series data.

### 2.2 Enhanced QPCA Algorithm

Unlike other methods such as those in the time domain, frequency domain, and time-frequency domain, the QPCA is a prevalent signal decomposition method. It generates features that are highly representative with minimal redundancy of information and can handle signals that are non-stationary and non-linear in nature [33]. Originally formulated for color image recognition, QPCA restricts the quaternion representation to pure quaternions due to the limitation of having only three color channels [31]. However, when applied to multichannel EEG processing, the traditional QPCA algorithm is enhanced with a full quaternion representation to enrich the mutual relationship information across channels. Meanwhile, to reduce computational load, the proposed algorithm is applied to the features of segmented signals, rather than to the raw data. In addition, to enhance the generalization power for classification, the features for classification are selected based on the principal components of the singular matrix. The enhanced QPCA algorithm for multichannel EEG signal processing is outlined in Algorithm 1 as follows:

**Algorithm 1: QPCA for Multichannel EEG Signal Processing**

1) For each subject $l$, four EEG channels, denoted as $Ch_1$, $Ch_2$, $Ch_3$, and $Ch_4$, are selected. $Ch_1(t)$, $Ch_2(t)$, $Ch_3(t)$, and $Ch_4(t)$ are real value measured at sampling point $t$ of the EEG signal under analysis. Let $T_w$ denote the length of whole time duration of the EEG signals under analysis, and $f_w$ is the sampling frequency. Then the total number of sampling points is $N_w = T_w \times f_w$.

2) To reduce the overall computational load, the EEG signals are segmented into short intervals (duration denoted as $T_s$) and the relative band power in these segmented signals are computed. Let $N_s = T_w/T_s$ be the number of segments. The relative band power of the delta band, theta band, alpha band, and beta band are denoted as $R_\sigma$, $R_\theta$, $R_\alpha$, and $R_\beta$. They are defined as the spectral power of the specific frequency band ($BP_\sigma$, $BP_\theta$, $BP_\alpha$, and $BP_\beta$) divided by the total spectral power ($BP$) of full frequency spectrum (1-30Hz) as shown in (6). Hence, the EEG signal is transformed into a vector of relative band power, each with length of $N_s$ per channel:

$$R_\alpha = BP_\alpha/BP, \text{ and similar for } R_\theta, R_\alpha, \text{ and } R_\beta \quad (6)$$

3) To analyze the characteristic of alpha band, the relative band power of alpha band, $R_\alpha^{Ch_i}, i \in (1,2,3,4)$ of these four EEG channels are used to construct a quaternion vectors of length $N_s$ as in (7), where $q_l \in \mathbb{H}^{1 \times N_s}$ and $n \in (1,\dots,N_s)$. Similar procedure is used in for the analysis of other frequency bands.

$$q_l(n) = R_\alpha^{Ch_1}(n) + R_\alpha^{Ch_2}(n)i + R_\alpha^{Ch_3}(n)j + R_\alpha^{Ch_4}(n)k \quad (7)$$

4) Let the training set contains $m$ subjects. Therefore $m$ quaternion vectors in alpha band $q_l \in \mathbb{H}^{1 \times N_s}, l = 1,\dots,m$, are constructed. The mean vector is defined as:

$$E(q_l) = \frac{1}{m}\sum_{l=1}^{m} q_l \quad (8)$$

The difference between each quaternion vector and the mean vector is denoted as:

$$\tilde{q}_l = q_l - E(q_l). \quad (9)$$

Then, a quaternion matrix $\tilde{Q} = (\tilde{q_1}^T,\dots,\tilde{q_m}^T)^T \in \mathbb{H}^{m \times N_s}$ is formed by concatenating all the centered quaternion vectors.

5) Compute the covariance matrix of $\tilde{Q}$ as:

$$C = \frac{1}{n}\tilde{Q}\tilde{Q}^H. \quad (10)$$

6) By SVD, the covariance matrix is decomposed as:

$$C = U\begin{pmatrix} \Sigma_r & 0 \\ 0 & 0 \end{pmatrix} U^H \quad (11)$$

$U \in \mathbb{H}^{N_s \times N_s}$ is the singular matrix, whose column vectors are eigenvectors of the covariance matrix $C$. The principle components are defined as the first $p$ eigenvectors in which the accumulated sum of corresponding eigenvalues exceeds a preset threshold. The eigenvectors space is denoted by $U_p \in \mathbb{H}^{N_s \times p}$.

7) The training feature is calculated by:

$$Y = \tilde{Q}U_p \quad (12)$$

Each row of $Y \in \mathbb{H}^{m \times p}$ is the training feature corresponding to each subject.

8) For testing each subject, follow Steps 1 to 3 to represent the 4 selected channels EEG signals by quaternion vectors $z_l \in \mathbb{H}^{1 \times N_s}$ and deducting the mean vectors $\tilde{z}_l = z_l - E(q_l)$. The testing feature $F$ can then be expressed as:

$$F = \tilde{z}_l U_p \quad (13)$$

9) As the training feature and testing feature are quaternion vectors, projection scheme is needed to transform the feature vector from quaternion domain into real domain for SVM classification. Let $q$, denoted in (1), as one of the quaternion value in the feature vector. The mean projection method is employed here as $\bar{q} = (x + y + w + z)/4$. Hence, the quaternion feature matrix is transformed into real valued feature matrix $F_{Real}$ and fitted to SVM classifier for detection.

$$F_{Real} = \bar{F} \quad (14)$$

## 2.3 Simulations

To assess the performance of the proposed QPCA algorithm in multichannel EEG processing, a series of simulations have been designed and implemented. A public EEG dataset for AD detection has been utilized. For the purpose of real-time analysis, the data dimension is significantly reduced through band power feature extraction. The enhanced QPCA algorithm is then applied, generating classification features for AD detection via an SVM classifier. General performance metrics such as accuracy, sensitivity, and specificity are employed for evaluation. Extensive simulations are conducted for all permutations of any 4-channel combinations of EEG signals. Based on these simulation results, the properties of QPCA as a measure for brain connectivity analysis are presented. Meanwhile, the most prominent channels and regions of interest in the brain for AD classification are identified and analyzed, referencing existing track records. Finally, the selection of several critical parameters in the enhanced QPCA algorithm, including the segmentation interval, the projection method, and the number of principal components, are further studied to optimize the performance. All the simulations were performed on a computer with the following specifications: Intel Core i7-10700K CPU @ 2.90GHz, 16 GB DDR4 RAM, and Windows 10 Pro 64-bit operating system.

### 2.3.1 Dataset

The dataset, sourced from Ziaeian Hospital in Tehran, Iran [34], was used for the simulation. This dataset contains EEG signals from 11 participants. The records of two other participants were excluded due to incomplete information. Among the selected subjects, five had memory complaints. They were diagnosed through neurological examination as either normal aging (non-AD) or suffering from mild AD. The EEG data were recorded using 19 mono-polar channels in the standard 10/20 system [35], as shown in Fig. 1. The data was preprocessed using EEGLAB [36] following Makoto's pipeline, which includes several noise filtering steps: (i) high-pass filtering at 1 Hz to remove slow drifts and low-frequency noise, (ii) notch filtering to eliminate line noise at 50/60 Hz, (iii) artifact subspace reconstruction (ASR) to remove transient artifacts, (iv) independent component analysis (ICA) to identify and remove components related to eye blinks, muscle activity, and other non-neural artifacts, and (v) low-pass filtering at 40 Hz to remove high-frequency noise. For each participant, EEG data were recorded during the presentation of an auditory stimulant. Each session consisted of six trials of 40-second stimuli, interleaved by 20-second rest intervals in silence. The entire task resulted in 340 seconds of EEG signal.

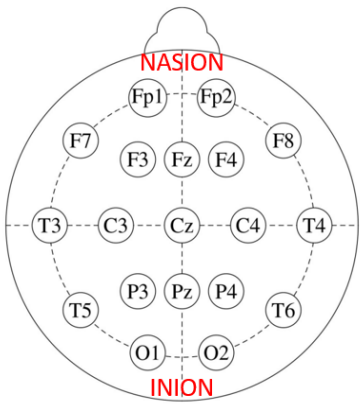

**Fig.1. Standard 10/20 EEG setup that consists of 19 channels.**

### 2.3.2 Simulation Procedures

From the EEG dataset, the six sessions during auditory stimulation for each participant are used for classification. The first five sessions are grouped as training samples, and the remaining one is used as the testing sample. This results in a total of 55 training samples and 11 testing samples. Each sample includes EEG recordings from 19 channels. As the stimuli last for 40 seconds at a 250 Hz sampling frequency, there are 10,000 sampling points for each channel. In primary experiments for performance evaluation, the length of each segment is set to one second, and the signal is divided into 40 segments. This choice is based on a systematic review by Cassani et al. of EEG studies for AD diagnosis [41], which indicated that the most common EEG epoch duration used in such studies was 1-2 seconds. However, to optimize the performance of our classification model, we conducted an additional experiment to compare the performance for different segment lengths. The spectral power of the data is estimated in power spectral method via Matlab BANDPOWER function. The frequency range is defined as 1Hz to 30 Hz, which contains the four major frequency bands: the alpha (1~4Hz), theta (4~8Hz), beta (8~13Hz), and delta (13~30Hz) bands. As a result, the segmented signal for each channel is transformed into a data vector with only 40 components of relative band power.

For each permutation of the 4-channel combinations, the enhanced QPCA algorithm is applied to the training dataset. Initially, a full quaternion representation of the selected four-channel data vector is constructed for each training sample. Then, a quaternion data matrix is formed by concatenating all the centered quaternion vectors, and its dimension equals to 55 (rows of training samples) x 40 (columns of relative band power). Using the Matlab QTFM toolbox, developed by S. Sangwine and N.L. Bihan [37], the covariance of the data matrix is computed and decomposed by quaternion SVD. The leading column vectors of the singular matrix act as the principal components. For evaluation purpose, the leading 20 principal components will be simulated separately to search for the highest accuracy. The MEAN projection methods are applied, converting the principal components from quaternion value to real value for binary classification of AD by the SVM classifier. The SVM classifier is implemented using the Matlab FITCSVM function, and the kernel function is set to the LINEAR function. Finally, accuracy, sensitivity, and specificity are used as statistical measures to evaluate the performance of the classification. The mathematical expressions of these statistical measures are given by:

$$\text{Accuracy (ACC)} = \frac{TP+TN}{TP+TN+FP+FN} \quad (15)$$

$$\text{Sensitivity (SEN)} = \frac{TP}{TP+FN} \quad (16)$$

$$\text{Specificity (SPE)} = \frac{TN}{TN+FP} \quad (17)$$

, where true positive (TP) is a test result that correctly indicates the presence of Alzheimer's disease (AD). In other words, TP is the number of AD patients who are correctly classified as having AD. True negative (TN) is a test result that correctly indicates the absence of a condition or characteristic. False positive (FP) is a test result which wrongly indicates that a particular condition or attribute is present. False negative (FN) is a test result which wrongly indicates that a particular condition or attribute is absent. The overall operation flow of the experiment is shown in Fig. 2. Furthermore, a 10-fold cross-validation on the existing dataset is conducted to validate the generalization of the QPCA algorithm. The cross-validation method is implemented using the Matlab *CROSSVAL* function with k=10. The experiments will be repeated 1000 times to ensure the robustness and reliability of the results.

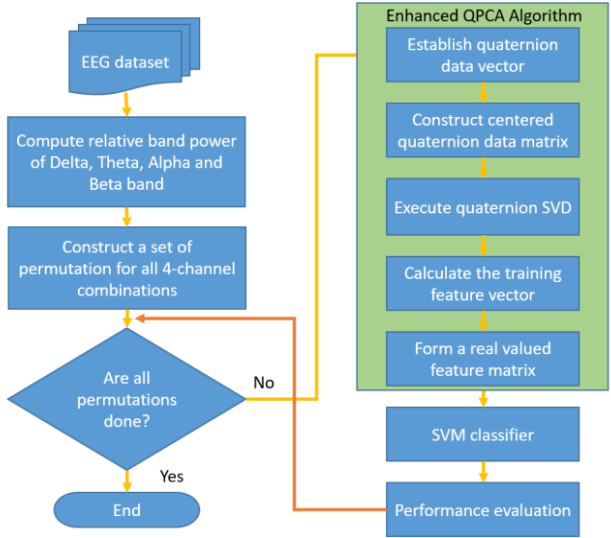

**Fig. 2. Flowchart of the simulation procedures.**

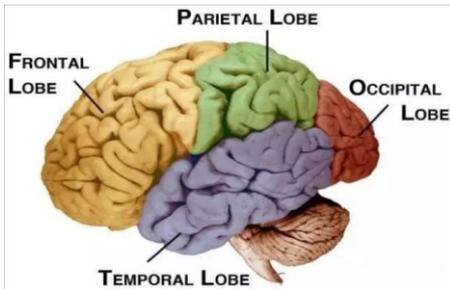

**Fig. 3. Diagrammatic representation of the major parts of the brain (lateral view).**

### 2.3.3 Brain Connectivity Analysis

Investigation of the relationship between EEG channels and early AD remains a hot topic. Some researchers have proposed selecting channels based on specific ranking methods. For instance, Puri *et al.* employed a subband-based energy to entropy ratio for channel ranking [38], while Perez-Valero *et al.* recently selected channels based on the statistical analysis of relative power features and complexity features [39]. However, the performance of these methods is data-dependent and lacks generalization. On the other hand, Cicalese *et al.* reported that an increasing number of researchers are determining the EEG channels in AD-related regions based on medical evidence [40]. As shown in Fig. 3, the right prefrontal area and left parietal area have been identified as being associated with AD-linked cognitive decline. These regions are implicated in multiple cognitive processes, such as working memory, attention, and executive function, and their dysfunction may cause the cognitive deficits observed in AD.

In the simulation, all permutations of any 4-channel combinations are implemented to conduct a comprehensive search for the most critical channels or brain regions in AD classification. Given that there are 19 channels in the EEG dataset, there are $C_4^{19} = 3876$ trials in QPCA analysis. However, due to the noncommutative property in the quaternion domain, the permutation of these four channels should be considered as different inputs. Therefore, the complete search consists of 93024 trials. Compared with traditional functional connectivity measures, which only have 171 combinations of pairwise channels for 19 channels, the quaternion representation of brain connectivity can be used to explore high-dimensional correlations among channels and to generate a more comprehensive description of brain dynamics. In traditional functional connectivity measures, a 2D connectivity matrix is constructed to illustrate the measure values for different pairs of channels. However, it is challenging to present the dynamic functional connectivity in QPCA, which is a 4D matrix with leading principal components as the measure value. To illustrate such high-dimensional connectivity relations, the data dimension should be reduced, and only three channels are used to construct the quaternion. The same simulation procedure is executed, and only the first principal component is used to compute the measure value. Consequently, a 3D brain connectivity matrix is formed to illustrate the brain dynamics under different frequency bands of EEG. To compute the discrimination power of the QPCA measure value between the normal subjects (NS) and AD patients, the interclass distance (*Dist*) is calculated as follows:

$$Dist = \left| \frac{1}{N_{NS}} \sum PC_{fb}^{NS} - \frac{1}{N_{AD}} \sum PC_{fb}^{AD} \right|, fb \in (\delta, \theta, \alpha, \beta) \quad (18)$$

, where $PC_{fb}^{NS}$ and $PC_{fb}^{AD}$ are the measure value in different frequency bands based on principal component of NS and AD. $N_{NS}$ and $N_{AD}$ are the number of samples of NS and AD.

### 2.3.4 Selection of QPCA Parameters

In the QPCA algorithm, there are three critical parameters: segmentation interval, projection method, and number of principal components. The selection of these parameters can significantly affect the discrimination power of features and the overall performance in AD classification. Therefore, their settings need to be further rationalized and optimized for better performance. To reduce the computational load, the raw EEG signals are segmented into non-overlapping short intervals. The duration of these intervals can affect the temporal resolution. If the duration is too short, real-time implementation becomes difficult. However, if the duration is too long, there can be a loss of temporal information. Therefore, selecting a suitable interval duration is critical. In this study, twelve different durations of segmentation intervals, as shown in Table 1, are tested and evaluated based on classification performance.

The projection method in the QPCA algorithm is used to convert the feature value from the quaternion domain into the real domain for SVM classification. This conversion is necessary because the SVM classifier operates in the real domain. The advantages of this approach include ensuring compatibility with the SVM classifier, preserving the essential information captured by QPCA, potentially enhancing classification performance, and facilitating standardization and comparison with other studies. Apart from the mean projection method, several other projection methods are implemented for comparison. The absolute projection method calculates the mean of the absolute value of each quaternion component. The norm projection method computes the square root of the sum of the squares of each quaternion component. The phase projection method extracts the phase angle of the quaternion. These methods are summarized in Table 2. As only a few leading principal components are employed for feature extraction, the selection of the number of principal components is very important. Simulations based on different numbers of principal components (from 1 to 40) will be conducted to investigate their performance in AD classification. Additionally, the comparison results will be analyzed to develop a better selection scheme. All of the above simulations are conducted on the same channel set, which has the highest accuracy in the alpha band. The performance metric is solely dependent on classification accuracy.

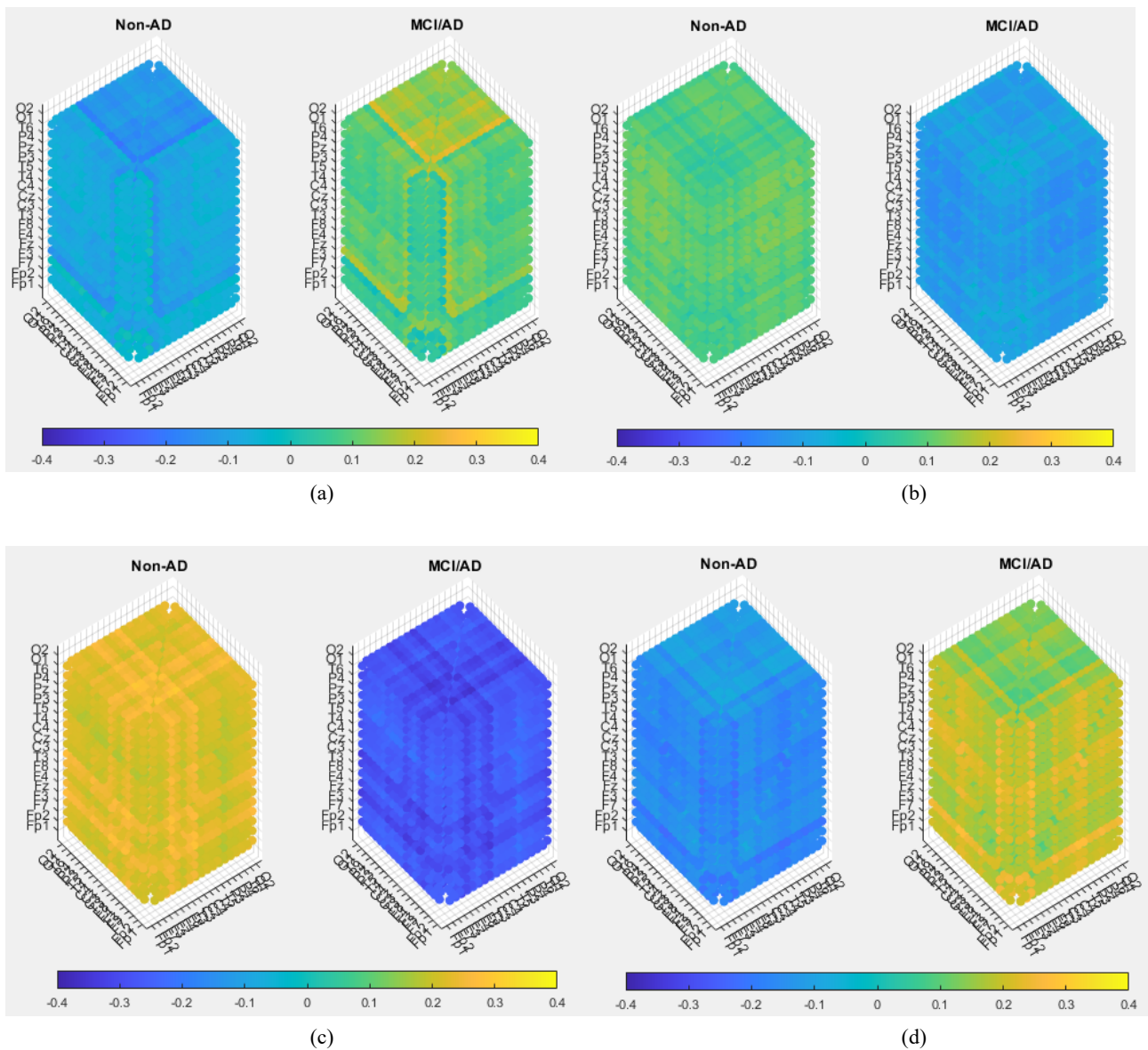

Fig. 4. 3D connectivity matrix for AD patients and normal healthy subjects (Non-AD) in different frequency bands, (a) delta band, (b) theta band, (c) alpha band, and (d) beta band. The colorbar depicts the amplitude of the QPCA measure value. The theta band and the alpha band show a decrease in amplitude for normal subjects compared with AD patients, but the delta band and the beta band show an increase. These results aligned with existing research findings.

**Table 1.  Segmentation interval settings.**

|  | Simulation Settings | | | | | |
|---|---|---|---|---|---|---|
| Case No. | 1 | 2 | 3 | 4 | 5 | 6 |
| Segment Interval (sec) | 0.1 | 0.2 | 0.4 | 0.5 | 0.8 | 1 |
| Length of Data Vector | 400 | 200 | 100 | 80 | 50 | 40 |
| Case No. | 7 | 8 | 9 | 10 | 11 | 12 |
| Segment Interval (sec) | 1.25 | 2 | 2.5 | 4 | 5 | 10 |
| Length of Data Vector | 32 | 20 | 16 | 10 | 8 | 4 |

**Table 2. Summary of projection methods for simulations**

|  | Projection Method | Implementation $q = w + x\boldsymbol{i} + y\boldsymbol{j} + z\boldsymbol{k}$ |
|---|---|---|
| 1 | Mean | $\bar{q} = (w + x + y + z)/4$ |
| 2 | Absolute | $|\bar{q}| = (|w| + |x| + |y| + |z|)/4$ |
| 3 | Norm | $\|q\| = \sqrt{w^2 + x^2 + y^2 + z^2}$ |
| 4 | Phase | $\theta = \tan^{-1}\sqrt{x^2 + y^2 + z^2}/w$ |

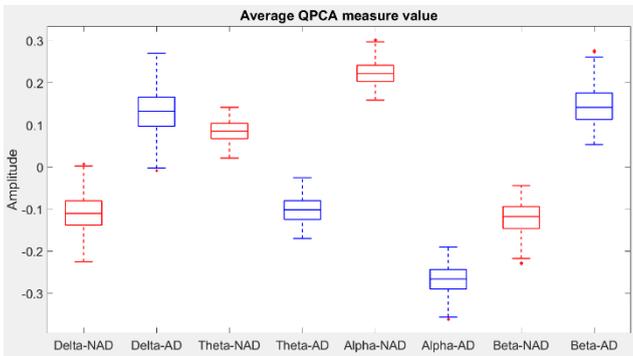

**Fig. 5. Boxplot showing distributions of average QPCA measure value under different frequency bands for both AD patients and normal subjects.**

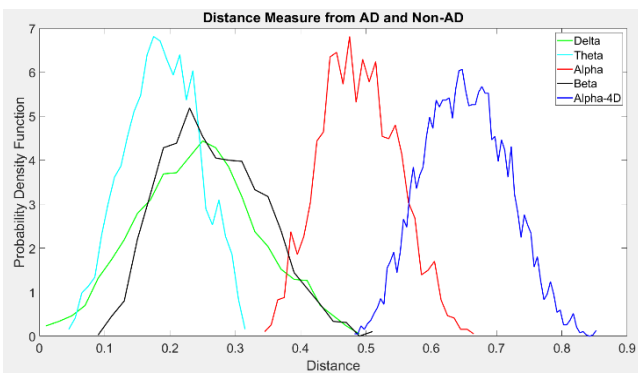

**Fig. 6. Histogram plot of distance measure from AD and Non-AD subjects based on the measure value of QPCA in different frequency bands.**

## 3. Results

### 3.1 Connectivity Measure

In Fig. 4, the 3D connectivity matrices for AD patients and normal healthy subjects in different frequency bands are shown. The color depicts the amplitude of the measure value. The statistics of these distributions are further computed and plotted in Fig. 5. It can be seen that the variation of the measure value under different frequency bands is different. The theta band and the alpha band show a decrease in amplitude for normal subjects compared with AD patients, but the delta band and the beta band show an increase. Referring to other related research findings, a decline of synchronization in alpha bands is mostly reported for AD patients [41], and an increase of coupling measure is found in the delta band and beta band for MCI patients [42]. However, it is hard to directly interpret these variations as the synchrony of EEG signals may be significantly affected by brain events other than changes of synchrony due to AD [43]. Furthermore, the probability distribution functions of the average absolute difference between normal subjects and AD patients under different frequency bands are computed to evaluate the discrimination power of these measure values. As shown in Fig. 6, the alpha band has the largest distance compared with the other frequency bands, hence its QPCA features can lead to higher accuracy in AD classification. In addition, the measure value based on a full representation (4D) QPCA in the alpha band is simulated for comparison. The distance measure of the alpha band in 4D outperforms the alpha band in pure quaternion representation (3D) by almost 30% as shown in Fig. 6. The larger discrimination power of the full quaternion representation is mainly caused by the supplement of high-dimensional correlation information among EEG channels.

### 3.2 Prominent Channels and Regions

After simulations for all permutations of 4-channel combinations, the performance of AD classification for each permutation in different frequency bands is evaluated. As there are 24 permutations for each combination, the classification performance of each combination is computed as the average of the results over all its permutations. As shown in Fig. 7, the classification performances for different combinations in the alpha band are similar, with the highest accuracy being 95%. Owing to the highest discriminative power of QPCA measure values in the alpha band compared with other frequency bands, almost all combinations have achieved good performance with an average accuracy of 90%. For the other frequency bands, the classification performance levels are more diverse, but some combinations still achieved 95% accuracy in the delta band. Furthermore, all the performance metrics in the alpha band is plotted in Fig. 8. The average accuracy, sensitivity and specificity are 90%, 88%, and 80% respectively. To address the concern regarding the generalization of the QPCA algorithm due to the limited dataset size, we conducted a 10-fold cross-validation on the existing dataset. The mean evaluation score obtained in alpha band with the combination (F8, T7, T8, P4) was 0.0942, indicating that the average accuracy of the model is approximately 90.58%. This comprehensive evaluation helps mitigate the limitations of the small sample size and provides a reliable estimate of the model's generalization performance.

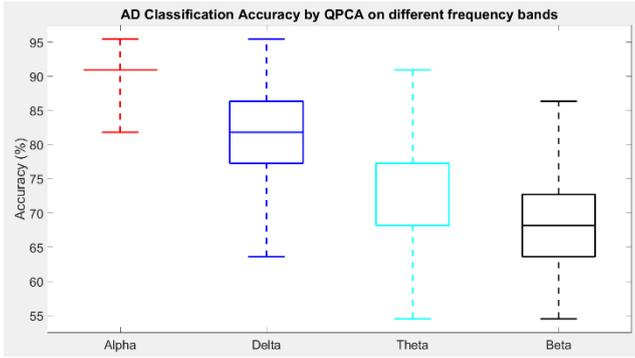

**Figure 7. Boxplot showing distributions of average accuracy under different frequency bands in AD classification.**

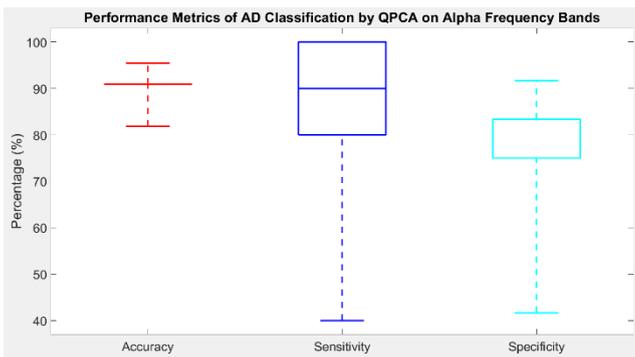

**Figure 8. Boxplot showing distributions of all performance metrics on alpha frequency bands in AD classification.**

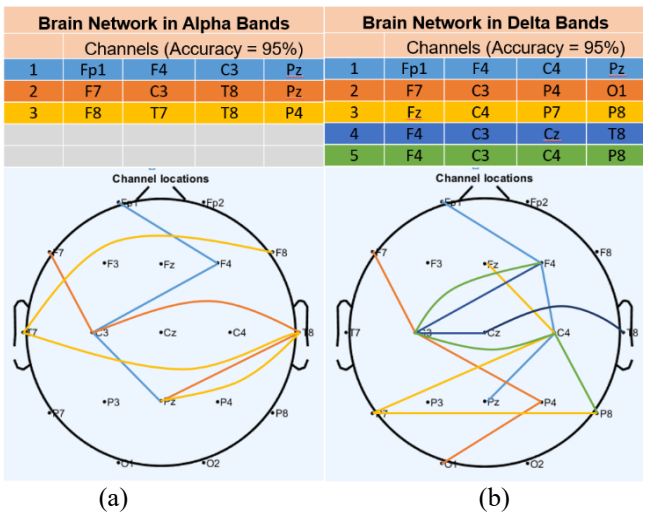

**Fig. 9. Brain network diagrams highlighting the highest AD classification accuracy (95%) in (a) alpha band, and (b) delta band.**

The performance metrics of the 24 permutations for the combination (F8, T7, T8, P4) with highest classification accuracy in the alpha band is shown in Table 3. Some permutations have attained 100% accuracy rate, while some are lower. Some specific permutations could achieve higher accuracy because they optimize the representation of the spatial and functional relationships between the selected channels. Quaternions can capture the complex interactions and dependencies between multiple channels [29-32]. The order in which channels are embedded into the quaternion can influence how these interactions are represented. Furthermore, certain permutations may align the channels in a way that maximizes the correlation between them, making it easier for the QPCA to extract meaningful features. Our results shown that QPCA is an effective feature extraction tool in multichannel EEG processing.

**Table 3. Performance metrics of the 24 permutations for the (F8, T7, T8, P4) combination with the highest average accuracy at 95%.**

| Permutation | ACC (%) | SEN (%) | SPE (%) |
|---|---|---|---|
| F8,T7,T8,P4 | 90.91 | 100 | 83.33 |
| F8,T7,P4,T8 | 100 | 100 | 100 |
| F8,T8,T7,P4 | 100 | 100 | 100 |
| F8,T8,P4,T7 | 90.91 | 100 | 83.33 |
| F8,P4,T7,T8 | 90.91 | 100 | 83.33 |
| F8,P4,T8,T7 | 100 | 100 | 100 |
| T7,F8,T8,P4 | 100 | 100 | 100 |
| T7,F8,P4,T8 | 90.91 | 100 | 83.33 |
| T7,T8,F8,P4 | 90.91 | 100 | 83.33 |
| T7,T8,P4,F8 | 100 | 100 | 100 |
| T7,P4,F8,T8 | 100 | 100 | 100 |
| T7,P4,T8,F8 | 90.91 | 100 | 83.33 |
| T8,F8,T7,P4 | 90.91 | 100 | 83.33 |
| T8,F8,P4,T7 | 100 | 100 | 100 |
| T8,T7,F8,P4 | 100 | 100 | 100 |
| T8,T7,P4,F8 | 90.91 | 100 | 83.33 |
| T8,P4,F8,T7 | 90.91 | 100 | 83.33 |
| T8,P4,T7,F8 | 100 | 100 | 100 |
| P4,F8,T7,T8 | 100 | 100 | 100 |
| P4,F8,T8,T7 | 90.91 | 100 | 83.33 |
| P4,T7,F8,T8 | 90.91 | 100 | 83.33 |
| P4,T7,T8,F8 | 100 | 100 | 100 |
| P4,T8,F8,T7 | 100 | 100 | 100 |
| P4,T8,T7,F8 | 90.91 | 100 | 83.33 |

In Fig. 9, the brain network with the highest classification accuracy, including the alpha band and delta band, are plotted topographically. The channel combinations with the highest accuracy reflect the most prominent channels or regions in AD classification. These channels include Fp1, F4, F7, Fz, F8, T7, C3, Cz, C4, T8, P7, Pz, P4, P8, and O1, and can be grouped into several brain regions such as frontal (Fp1, F4, Fz), temporal (F7, F8, T7, T8, P7, P8), parietal (C3, Cz, C4, Pz, P4), and occipital (O1). The simulation results show that the temporal lobe is one of the most prominent regions in AD classification, as the EEG signals are recorded during the presentation of the auditory stimulant. The auditory cortex in the brain, which is part of the temporal lobe, processes auditory information. The loss of functional connectivity in the temporal lobe of AD patients becomes more critical than other regions of the brain in AD classification [44]. In addition to the commonly associated regions such as the frontal lobe and parietal lobe in cognitive decline [45], our results reveal that QPCA features provide a consistent measure in AD classification.

### 3.3 QPCA Parameters Optimization

#### 3.3.1. Selection of Segmentation Interval

Based on different segmentation interval settings in Table 1, the AD classification simulations are implemented with one of the highest accuracy channels in the alpha band (F8, T7, T8, and P4). The simulation result is shown in Fig. 10. Different segmentation intervals yield different performances, and the highest accuracy is obtained when the segmentation interval is only 1 second. The simulation result provides quantitative evidence for the most common EEG epoch duration as stated in [41]. It also implies that the temporal information of the EEG signals is critical to the classification performance. Therefore, it is better to determine the most suitable interval before further analysis of EEG, as it is highly data-dependent.

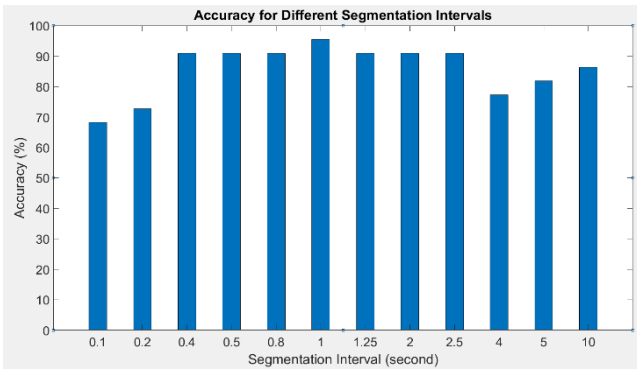

**Fig. 10. Accuracies for different segmentation intervals in AD classification.**

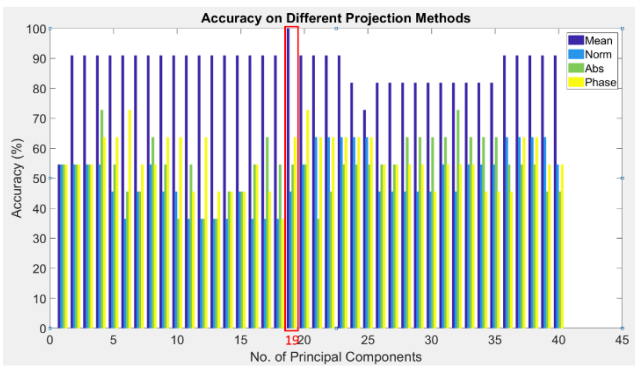

**Fig. 11. Accuracies for different projection methods.**

#### 3.3.2. Selection of Projection Methods

Fig. 11 denotes the performance in AD classification when applying different projection methods as shown in Table 2. The average and highest accuracy of the mean projection method equals 87% and 100% respectively, which outperforms the performance of other methods. This can be explained by the QPCA measure values of AD patients and healthy subjects. If the sign of the values is neglected as in the other projection methods, the inter-class distance will be reduced, and the classification performance will be weakened. On the other hand, the mean projection method considers the sign of the values and retains the original inter-class distance.

#### 3.3.3. Selection of Principal Components

The performance in AD classification when applying different numbers of principal components in feature extraction is shown in Fig. 12. The simulation is done on one of the highest accuracy permutation (F8, T7, P4, T8) in Table 3. The best performance, with an accuracy of 100%, occurs when the number of leading principal components is 19. However, it is difficult to identify the number of leading principal components in the general aspect. A common method is based on the eigenvalues of the decomposition, and the best performance will be attained when the sum of the eigenvalues of the leading principal components exceeds a threshold (e.g., 90%). In the simulation, there are three combinations in the alpha band with the highest accuracy and the sum of their eigenvalues are about 87%, 90%, and 91%. But, this value varies significantly in different frequency bands. For example, there are five combinations in the delta band with the highest accuracy and the sum of their eigenvalues are 80%, 74%, 84%, 90%, and 93%. Therefore, a better approach should be developed for the selection of the number of principal components in EEG analysis.

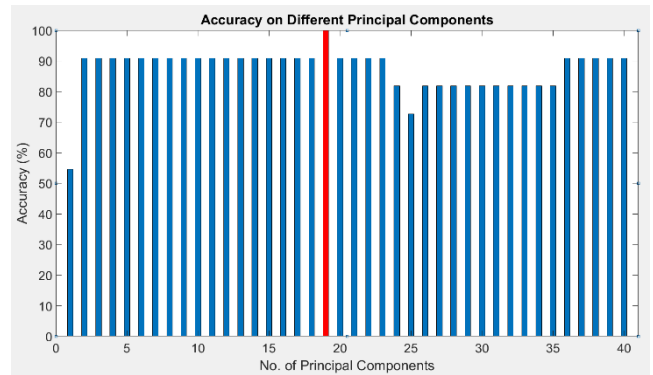

**Fig. 12. Accuracies for applying different number of principal components in feature extraction.**

### 4. Discussion

In this paper, the quaternion representation of multichannel EEG signals enhances the high-dimensional correlation among the channels and provides a more thorough description of brain connectivity in 4D. By integrating with the PCA approach, QPCA consolidates the dynamic variations of the functional connectivity among the channels over EEG recording and generates compact and effective features for brain disease classification. Furthermore, the 4D QPCA values can be used as a high-dimensional connectivity measure for dynamic functional connectivity analysis.

## 4.1. Comparison with Traditional Method and Relevant Studies

For comparison, a traditional PCA method was implemented by concatenating the selected four-channel feature vectors into one real vector and applying PCA to obtain the principal components. These real-valued principal components were used to construct the training feature matrix and were fitted to an SVM classifier for binary classification of AD. The accuracy of the proposed QPCA algorithm is 100%, while the traditional PCA method achieved an accuracy of 82%. The proposed QPCA algorithm outperforms the traditional PCA method, as quaternion representation and manipulation extract more cross-channel interrelations, resulting in classification features based on quaternion eigenvectors that have greater discriminative power than those derived from the traditional PCA method in AD detection. Table 4 provides a comparison of the performance of the proposed method in term of accuracy with those reported in related recent works. The proposed algorithm demonstrates superior performance in terms of AD classification accuracy, highlighting the effectiveness of the quaternion-based PCA and SVM machine learning approach.

**Table 4. Performance Comparison in AD Classification Based on EEG Dataset**

| Study | Methodology | ACC (%) | SEN (%) | SPE (%) |
|---|---|---|---|---|
| Wantanabe et al. (2024) [46] | Deep convolution neural network | 93.4 | Nil | Nil |
| Siuly et al. (2024) [47] | Statistical features and spectral power with LSTM network | 97.0 | 97.0 | 98.4 |
| Vicchietti et al. (2023) [48] | Quantiles graphs and wavelet coherence with SVM classifier | 100 | 100 | 100 |
| Puri et al. (2023) [49] | Orthogonal wavelet filter banks and fractal dimension with SVM classifier | 98.6 | 97.3 | 99.8 |
| Alessandrini et al. (2022) [50] | Robust-PCA and LSTM RNN | 97.9 | Nil | Nil |
| Safi et al. (2021) [51] | Hjorth parameters and DWT with KNN classifier | 97.6 | 95.4 | 98.8 |
| This Work | QPCA with SVM classifier | 100 | 100 | 100 |

## 4.2. 4D Connectivity Measure

Compared with a 2D connectivity measure, the 4D connectivity matrix contains more details in the interrelations among channels, therefore, it has more discriminative power and can separate different levels of neurodegenerative disease. However, the 4D connectivity matrix greatly enlarges the load of data manipulation. The average calculation time for the training phase was approximately 0.4 seconds and for the testing phase, it was approximately 0.02 second. The most computationally intensive part of QPCA is the QSVD operation, which involves a full complex quaternion-valued matrix decomposition. As only a few leading principal components from the decomposition are used for feature construction, a partial decomposition approach is proposed to reduce the computation load. Recently, a Lanczos-based method has been proposed to enhance the computational efficiency of partial SVD triplets of large-scale quaternion matrices [52]. This method iteratively approximates the leading singular values and vectors, significantly reducing the number of computations required compared to a full QSVD. The Lanczos-based QSVD method has demonstrated better performance compared to state-of-the-art methods, with practical implementations showing a dramatic reduction in computation time [53]. Therefore, by using Lanczos-based QSVD, QPCA will be more computationally efficient and suitable for real-time analysis of 4D connectivity matrix.

## 4.3. Prominent Channels and Regions

Although the prominent channels and regions of the brain identified by the QPCA simulations are very consistent with relevant research findings, these regions of interest are very dependent on the activity during EEG recording. In the simulation, the dataset is collected during the presentation of an auditory stimulant, therefore the EEG channels in the temporal lobe show greater differences in brain connectivity between AD patients and healthy subjects [54]. If the EEG signals are recorded under memory tests or while watching a video, the frontal and parietal lobes [40] or the occipital lobe [55] should be more determinant. Therefore, most EEG analyses for AD diagnosis and progression assessment are based on resting-state EEG recording. As the participant is not required to perform any specific task, EEG acquisition becomes simpler, more comfortable, and less stressful for the patient. However, event-related EEG recordings offer the opportunity to examine the effect of AD on specific brain circuits [41]. Depending on the stimulus presented during EEG recording, brain connectivity measures may be more or less discriminative for early AD detection, therefore a systematic exploration of different stimuli on pre-clinical AD diagnosis should be conducted.

## 4.4. Selection of Leading Principal Components

To further enhance the performance of QPCA, a better approach in selecting the number of leading principal components is critical. Many methods have been proposed for this selection. As the leading principal components reflect the largest variance of the data, a common approach is to select those components in which the cumulative sum of their corresponding eigenvalues exceeds a certain percentage (e.g., 90%) of the total sum, so that the selected components represent most of the information of the data. However, data variance does not directly relate to classification accuracy. This approach does not benefit classification because the information for various classes of objects has not been exploited. One promising method to enhance the QPCA algorithm for classification is the application of linear discriminant analysis, which finds a linear combination of features that best separates the data into various classes by maximizing the interclass separation and minimizing the intra-class separation of the feature vectors. It has been shown to be effective for improving the classification accuracy in many pattern recognition applications [56]. Instead of depending on the eigenvalues, a ratio of the interclass separation to the intra-class separation is defined, and those principal components with a large ratio value will be selected to construct features in AD classification.

While the current study demonstrates the potential of QPCA for AD classification using a small dataset, the need for larger datasets to validate the generalizability of the findings is acknowledged. Efforts are underway to collaborate with additional research institutions and access larger EEG datasets. Future work will focus on conducting extensive validation studies and exploring data augmentation techniques to enhance the robustness and applicability of the proposed method. To further enhance the feature extraction, the integration of QPCA with other EEG analysis techniques, such as wavelet transforms or time-frequency analysis, presents a promising direction for future research. In addition, its ability to identify patterns and anomalies could be harnessed for diagnosing other brain-related diseases and mental disorders. It could potentially discern different levels of pathology within a disease, offering a nuanced understanding of a patient's condition. This could lead to more personalized treatment plans, ultimately improving patient outcomes.

## 5. Conclusion

This paper introduces an improved QPCA algorithm for the AD classification and the analysis of brain connectivity. When compared to a 2D connectivity measure, the 4D connectivity matrix produced by the QPCA algorithm provides a more detailed representation of the interrelations among EEG channels and offers greater discriminative power in distinguishing different stages of neurodegenerative disease. The simulation reveals a decrease in connectivity in alpha bands and an increase in the delta and beta bands, aligning with existing research findings. The average accuracy for all combinations in alpha band reached at 95%, while some particular permutations achieved a perfect accuracy rate of 100%. These results demonstrate the high accuracy of AD classification using QPCA-based features. Furthermore, the temporal lobe emerges as one of the most significant regions in AD classification. Given that the EEG signals are recorded during the presentation of an auditory stimulant, the loss of functional connectivity in the temporal lobe of AD patients is more critical for AD classification than other brain regions. To further optimize the QPCA algorithm, we evaluated the selection of several key parameters, including the segmentation interval, the projection method, and the number of leading principal components, and proposed recommendations for future research. To our knowledge, this is the first application of the QPCA algorithm in AD classification and the first generation of a 4D connectivity matrix for brain connectivity analysis.

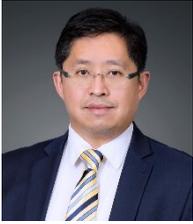
**Kevin Hung** received his Ph.D. in Electronic Engineering (concentration in biomedical engineering) from The Chinese University of Hong Kong. He is currently an Associate Professor and the Head of the Department of Electronic Engineering and Computer Science in the School of Science and Technology, Hong Kong Metropolitan University. His research interests include wearable technologies, mobile health, biological system modelling, engineering education, and quantum machine learning. He is currently serving as the Vice Chair of IEEE Hong Kong Section and Past Chair of the Electronics and Communications Section at IET Hong Kong.

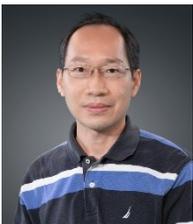
**Gary Man-Tat Man** is Research Assistant of the Department of Electronic Engineering and Computer Science in the School of Science and Technology at the Hong Kong Metropolitan University, Hong Kong. He received his B.Eng. and Ph.D. degrees in Electronic and Information Engineering from the Hong Kong Polytechnic University, Hong Kong, in 1990 and 1998, respectively. His current research interests include biomedical signal processing, eye tracking, quaternion representation and processing, and brain-related disease classification.

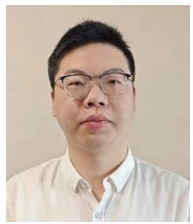
**Jincheng Wang** received his B. Sc. degree in Computer Engineering and M. Sc. degree in Scientific Research from the Hong Kong Metropolitan University (HKMU), Hong Kong, in 2021 and 2023, respectively.